\documentclass{article}
\usepackage{spconf,amsmath,amssymb,graphicx}
\usepackage{epstopdf}
\usepackage{enumitem}
\usepackage{array}
\usepackage{epstopdf}
\usepackage{cite}
\usepackage{hyperref}
\usepackage{flushend}
\usepackage{booktabs}
\usepackage[numbers,sort&compress]{natbib}
\newcommand{\vectornorm}[1]{\left\|#1\right\|}

\title{ANGULAR RESOLUTION LIMIT FOR DETERMINISTIC CORRELATED SOURCES}

\name{Xin Zhang$^*$, Mohammed Nabil El Korso$^{**}$ and Marius Pesavento$^*$}
\address{$^*$Communication Systems Group, Technische Universit\"{a}t Darmstadt, Darmstadt, Germany\\
$^{**}$ENS-Cachan / SATIE, Bagneux, France\\
Accepted, ICASSP'2013}

\begin{document}
\ninept
\onecolumn

\maketitle

\begin{abstract}
This paper is devoted to the analysis of the angular resolution limit (ARL), an important performance measure in the directions-of-arrival estimation theory. The main fruit of our endeavor takes the form of an explicit, analytical expression of this resolution limit, w.r.t. the angular parameters of interest between two closely spaced point sources in the far-field region. As by-products, closed-form expressions of the Cram\'er-Rao bound have been derived. Finally, with the aid of numerical tools, we confirm the validity of our derivation and provide a detailed discussion on several enlightening properties of the ARL revealed by our expression, with an emphasis on the impact of the signal correlation.
\end{abstract}

\section{Introduction}
\label{sec:intro}

As an important topic within the area of signal processing, far-field source localization in sensor array has found wide-ranging application in, among others, radar, radio astronomy and wireless communications \cite{KV96}. One common measure to evaluate the performance of this estimation problem is the resolvability of closely spaced signals, in terms of their parameters of interest. In this paper we investigate the minimum angular separation required under which two far-field point sources can still be correctly resolved.

To approach this problem, it is necessary to revive the concept of the resolution limit (RL), which will serve as the theoretical cornerstone of this paper. The RL is commonly defined as the minimum distance w.r.t. the parameter of interest (e.g., the directions-of-arrival (DOA) or the electrical angles, etc.), that allows distinguishing between two closely spaced sources \cite{S05,ShaMil05,EBRM10}. Till now there exist three approaches to describe the RL. The first rests on the analysis of the mean null spectrum \cite{C73}, the second on the detection theory \cite{ShaMil05,SM04,LN07}, and the third on the estimation theory, capitalizing on the Cram\'{e}r-Rao bound (CRB) \cite{L92,S05,EBRM11a}. A widely accepted criterion based on the third approach, proposed by Smith \cite{S05}, states that {\it two source signals are revolvable if the distance between the sources (w.r.t. the parameter of interest) is greater than the standard deviation of the distance estimation}. In this paper we consider the RL in the Smith's sense, due to the following advantages over competing approaches: The Smith criterion \textit{i}) takes the coupling between the parameters into account and thus is preferable to other criteria of the same category, e.g., the one proposed in \cite{L92,L94,D98}; \textit{ii}) enjoys generality unlike, e.g., the mean null spectrum approach which is designed for a specific high-resolution algorithm; \textit{iii}) is closely related to the detection theory approach, as recently revealed in \cite{EBRM10}.

This paper investigates the analysis of the RL for two closely spaced correlated deterministic sources. The RL has recently received an increasing interest especially after the publication \cite{S05} which received the IEEE best paper award. Prior works on the RL consider, on the one hand some specific criteria as the RL based on hypothesis tests \cite{AW08,ShaMil05,zhxtc6,zhxtc7}, or are tailored to specific estimation procedures as, e.g., the MUSIC algorithm in \cite{AD08}. On the other hand prior works based on the Smith criterion either contain non-closed form expressions that require numerical evaluation, as in \cite{AJS09,LN07} or are based on specific ideal assumptions (e.g., one known DOA \cite{zhxtc8,S05}, uncorrelated sources \cite{AD08}, ULA case \cite{S05,zhxtc8,EBRM11a}, non-time-varying sources \cite{EBRM11a}, etc.) In our work, we propose to derive an analytical expression for the angular resolution limit (ARL)\footnote{The so-called ARL characterizes the RL when we consider the angular parameters as the unknown parameters of interest.}, denoted by $\delta$, between two closely spaced, time-varying (both in amplitude and phase) far-field point sources impinging on non-uniform linear array, which, to the best of our knowledge, is till now absent in the current literature. As by-products, closed-form expressions of CRB w.r.t. the relevant parameters are provided to facilitate the derivation of the ARL. Furthermore, our expression, by virtue of its concise form highlights the respective effects of various system parameters on the ARL $\delta$. The analysis of expressions for different cases of correlated and uncorrelated sources, reveals a number of enlightening properties pertinent to the ARL's behavior, while being also computationally efficient, avoiding the difficulties associated with the numerical solution of non-linear equations.

The following notation will be used throughout this paper: $(\cdot)^{H}$, $(\cdot)^{T}$ denote the conjugate transpose and the transpose of a matrix, respectively. $\mbox{tr}\{\cdot\}$ and $\mbox{vec}\{\cdot\}$ denote the trace and the vectorization of a matrix, respectively. $\Re\{\cdot\}$ and $\Im\{\cdot\}$ denote the real and imaginary part respectively. $\vectornorm{\cdot}$ denotes the norm of a vector, $\otimes$ denotes the Kronecker product, whereas $E\{\cdot\}$ denotes expectation.

\section{Model Setup}
Consider a linear, possibly non-uniform, array comprising $M$ sensors that receives two narrowband time-varying far-field sources $s_{1}(t)$ and $s_{2}(t)$, the directions-of-arrival of which are $\theta_{1}$ and $\theta_{2}$, respectively. Then the received signal at the \textit{m}-th sensor can be expressed as \cite{KV96}:
\begin{equation}\label{wc1}
\begin{aligned}
&x_{m}(t)=\sum_{i=1}^{2}s_{i}(t)e^{jkd_m\sin(\theta_{i})}+n_{m}(t),\quad t=1,\dots,N\\
&\mbox{and} \quad m=1,\dots,M,
\end{aligned}
\end{equation}where the sources are modeled by\footnote{Note that this is a commonly used signal model in communication systems (cf. \cite{God97,LSZ96}).} $s_{i}(t)=a_{i}(t)e^{j(2\pi f_0+\pi_i(t))},\ i=1,2$ in which $a_i(t)$ denotes the time-varying non-zero real amplitude, $f_0$ denotes the carrier frequency, $\pi_i(t)$ denotes the time-varying phase, $d_{m}$ denotes the spacing between the first sensor (which is chosen as the so-called reference sensor, i.e., $d_{1}=0$) and the \textit{m}-th sensor, $k=\frac{2\pi}{\lambda}$ is the wave number (with $\lambda$ denoting the wave length), $n_{m}(t)$ denotes the additive noise at the \textit{m}-th sensor, and $N$ is the number of snapshots.

Fore mathematical convenience, we define $\nu_{i}=k\sin(\theta_{i}),\ i=1,2$ as our parameters of interest. Changing (\ref{wc1}) into the vector form, one obtains:
\begin{equation}
\boldsymbol{x}(t)=\boldsymbol{A}\boldsymbol{s}(t)+\boldsymbol{n}(t),
\end{equation}where $\boldsymbol{x}(t)=\left[x_{1}(t),\dots, x_{M}(t)\right]^{T}$,
$\boldsymbol{s}(t)=\left[s_{1}(t), s_{2}(t)\right]^{T}$, $\boldsymbol{n}(t)=\left[n_{1}(t),\dots, n_{M}(t)\right]^{T}$, and $\boldsymbol{A}=\left[\boldsymbol{a}(\nu_{1}), \boldsymbol{a}(\nu_{2})
\right]$. The steering vectors are defined as $\boldsymbol{a}(\nu_{i})=[e^{j\nu_{i}d_1},\dots,e^{j\nu_{i}d_M}]^{T}, \ i=1,2$. Furthermore, define the correlation factor $\rho$ between the two signals as \cite{zhxtc3}
\begin{equation}\label{c2}
\rho=\frac{\boldsymbol{s}_{1}^{H}\boldsymbol{s}_{2}}{\vectornorm{\boldsymbol{s}_{1}}\vectornorm{\boldsymbol{s}_{2}}}
\end{equation}where $\boldsymbol{s}_{i}=[s_{i}(1),\dots,s_{i}(N)]^{T},\ i=1,2$ are signal vectors.

The following assumptions are made in the remaining of this chapter:
\begin{description}
\item[A1] The sensor noise follows a complex circular white Gaussian distributed, both spatially and temporally, with zero-mean
and unknown noise variance $\sigma^{2}$.

\item[A2] The source signals are assumed to be deterministic; and the separation of the sources is small.

\item[A3] The unknown parameter vector is $\boldsymbol{\xi}=\left[\nu_{1}, \nu_{2}, \sigma^{2}\right]^{T}$. Thus, for given $\xi$, the joint probability density function of the observation $\boldsymbol{\chi}=[\boldsymbol{x}^{T}(1),\dots,\boldsymbol{x}^{T}(N)]^{T}$ can be written as $p(\boldsymbol{\chi}\mid\boldsymbol{\xi})=\frac{1}{\pi^{MN}|\boldsymbol{R}|}\exp\left(-(\boldsymbol{\chi}-\boldsymbol{\mu})^H
    \boldsymbol{R}^{-1}(\boldsymbol{\chi}-\boldsymbol{\mu})\right)$, where $\boldsymbol{R}=\sigma^2\boldsymbol{I}_{MN}$ and $\boldsymbol{\mu}=\left[
    (\boldsymbol{A}\boldsymbol{s}(1))^T,\dots,(\boldsymbol{A}\boldsymbol{s}(N))^T\right]^T$.
\end{description}

\section{Derivation of $\delta$}
The derivation of the ARL $\delta$ can be divided into three steps. The first step involves the derivation of the CRBs w.r.t. the relevant parameters. The second builds on this result and simplifies the implicit function based on the Smith criterion, the root of which yields $\delta$. The last step is to solve the function corresponding to different values of $\rho$, leading to the final expression for the ARL.

\subsection{CRB Derivation}
The CRB of the unknown parameters ($\nu_{1}$ and $\nu_{2}$) is obtained as the analytical inverse of the Fisher information matrix (FIM) for $\boldsymbol{\xi}$ (denoted by $\boldsymbol{\mathcal{I}}$). Under Gaussian noise, the elements of $\boldsymbol{\mathcal{I}}$ can be calculated using the following expression \cite{SM05}:
\begin{equation}\label{n1}
\left[\boldsymbol{\mathcal{I}}\right]_{i,j}=\mbox{tr}\left\{
\boldsymbol{R}^{-1}
\frac{\partial\boldsymbol{R}}{\partial\left[\boldsymbol{\xi}\right]_{i}}
\boldsymbol{R}^{-1}
\frac{\partial\boldsymbol{R}}{\partial\left[\boldsymbol{\xi}\right]_{j}}\right\}
+2\Re\left\{
\frac{\partial\boldsymbol{\mu}^{H}}{\partial\left[\boldsymbol{\xi}\right]_{i}}
\boldsymbol{R}^{-1}
\frac{\partial\boldsymbol{\mu}}{\partial\left[\boldsymbol{\xi}\right]_{j}}
\right\}.
\end{equation}where $\left[\boldsymbol{\xi}\right]_{i}$ denotes the \textit{i}-th element of the parameter vector $\boldsymbol{\xi}$. Thus for our model, $\boldsymbol{\mathcal{I}}$ takes the following block-diagonal form:
\begin{align}\label{wz1}
\boldsymbol{\mathcal{I}}=\left[
\begin{array}{cc}
\boldsymbol{\mathcal{\bar{I}}} & \boldsymbol{0}\\
\boldsymbol{0}^{T} & \frac{MN}{\sigma^4}
\end{array}
\right],
\end{align}where
\begin{align}\label{wz11}
\boldsymbol{\mathcal{\bar{I}}}=\left[
\begin{array}{cc}
2N\alpha\mbox{SNR}_{1} & \frac{2}{\sigma^{2}}\Re\{\eta\}\\
\frac{2}{\sigma^{2}}\Re\{\eta\} & 2N\alpha\mbox{SNR}_{2}
\end{array}
\right],
\end{align}in which $\alpha=\sum_{m=1}^Md_m^2$, $\mbox{SNR}_{i}=\varepsilon_i^{2}/\sigma^{2},\ i=1,2$, where $\varepsilon_i=\sqrt{\sum_{t=1}^{N}a_i^2(t)/N},\ i=1,2$; and
 \begin{equation}\label{c3}
\eta=\boldsymbol{s}_{1}^{H}\boldsymbol{s}_{2}\sum^{M}_{m=1}d_m^{2}e^{-jd_m(\nu_{1}-\nu_{2})}=
\boldsymbol{s}_{1}^{H}\boldsymbol{s}_{2}\sum^{M}_{m=1}d_m^{2}e^{-jd_m\Delta},
\end{equation}where $\Delta=\nu_{1}-\nu_{2}$ denotes the spacing between $\nu_{1}$ and $\nu_{2}$. We assume in the following that $\nu_{1}>\nu_{2}$, hence $\Delta>0$.

By inverting the $2\times2$ matrix $\boldsymbol{\mathcal{\bar{I}}}$ we obtain the following expressions for the entries of the CRB matrix:
\begin{equation}\label{n2}
\mbox{CRB}(\nu_{1})\triangleq\left[\boldsymbol{\mathcal{\bar{I}}}^{-1}\right]_{1,1}=\frac{2N\alpha}{\Psi}\mbox{SNR}_{2},
\end{equation}
\begin{equation}\label{n3}
\mbox{CRB}(\nu_{2})\triangleq\left[\boldsymbol{\mathcal{\bar{I}}}^{-1}\right]_{2,2}=\frac{2N\alpha}{\Psi}\mbox{SNR}_{1},
\end{equation}and
\begin{equation}\label{n4}
\mbox{CRB}(\nu_{1},\nu_{2})\triangleq\left[\boldsymbol{\mathcal{\bar{I}}}^{-1}\right]_{1,2}=-\frac{2}{\sigma^{2}\Psi}\Re\{\eta\},
\end{equation}where $\Psi=4\alpha^{2}N^2\mbox{SNR}_{1}\cdot\mbox{SNR}_{2}-(4/\sigma^{4})\cdot\Re^{2}\{\eta\}$ is the determinant of $\boldsymbol{\mathcal{I}}$.

\subsection{Equating the ARL}
According to the Smith criterion, the ARL, $\delta$, is given as the angular spacing, $\Delta$, which is equal to the standard deviation of the estimate of $\Delta$. The latter, under mild conditions \cite{leh83}, can be approximated as $\sqrt{\mbox{CRB}(\Delta)}$, suggesting that $\delta$ can be obtained as the (positive) solution of the equation:
\begin{equation}\label{wc3}
\delta^{2}=\mbox{CRB}(\delta).
\end{equation}where $\mbox{CRB}(\delta)=\mbox{CRB}(\nu_{1})+\mbox{CRB}(\nu_{2})-2\mbox{CRB}(\nu_{1},\nu_{2})$ \cite{EBRM10}.

Substituting (\ref{n2})-(\ref{n4}) into (\ref{wc3}), the latter is transformed into:
\begin{equation}\label{n5}
\begin{aligned}
\delta^{2}&=\mbox{CRB}(\nu_{1})+\mbox{CRB}(\nu_{2})-2\mbox{CRB}(\nu_{1},\nu_{2})\\
&=\frac{2}{\Psi}(N\cdot\mbox{SNR}_{2}\alpha+N\cdot\mbox{SNR}_{1}\alpha+\frac{2}{\sigma^{2}}\Re\{\eta\}).
\end{aligned}
\end{equation}Substituting $\delta$ for $\Delta$ in identity (\ref{c3}) we observe that (\ref{n5}) is a highly non-linear equation in $\delta$. Hence, in order to obtain the solution of (\ref{n5}) w.r.t. $\delta$, and taking into account that $\delta$ is small, we resort to the first-order Taylor expansion of $\eta$ around $\delta=0 $\footnote{In asymptotic cases $\delta$ becomes very small and our approximation made here is tight, as will be proved by our simulation (cf. Fig.~\ref{fig:1}). This can be explained by the fact that the Maximum Likelihood estimator, and generally all high resolution estimators, have asymptotically an infinite resolution capability leading to $\delta\rightarrow0$ \cite{SN89,zhxtc9}.} to obtain:
\begin{equation}\label{c1}
\begin{aligned}
\eta&\approx\boldsymbol{s}_{1}^{H}\boldsymbol{s}_{2}\sum^{M}_{m=1}d_m^2(1-jd_m\delta)\\
&=\boldsymbol{s}_{1}^{H}\boldsymbol{s}_{2}\left(\sum^{M}_{m=1}d_m^2-j\delta\sum^{M}_{m=1}d_m^3\right)\\
&=\boldsymbol{s}_{1}^{H}\boldsymbol{s}_{2}(\alpha-j\delta\beta),
\end{aligned}
\end{equation}where $\beta=\sum_{m=1}^{M}d_m^3$. Combining (\ref{c1}) with (\ref{c2}), it follows that:
\begin{equation}\label{n6}
\begin{aligned}
\Re\{\eta\}&\approx\vectornorm{\boldsymbol{s}_{1}}\vectornorm{\boldsymbol{s}_{2}}\Re\{\rho(\alpha-j\delta\beta)\}\\
&=N\varepsilon_1\varepsilon_2(\bar{\rho}\alpha+
\tilde{\rho}\beta\delta),
\end{aligned}
\end{equation}where $\bar{\rho}$ and $\tilde{\rho}$ are defined as the real and imaginary part of $\rho$, respectively, i.e., $\bar{\rho}=\Re\{\rho\}$ and $\tilde{\rho}=\mathcal{I}\{\rho\}$.

Now we merge (\ref{n6}) into (\ref{n5}) and, after some mathematical manipulations, obtain the following quartic function of $\delta$:
\begin{equation}\label{n7}
D^{2}\delta^{4}+2CD\delta^{3}+(C^{2}-AB)\delta^{2}+D\delta+\frac{A+B}{2}+C=0,
\end{equation}
where $A$, $B$, $C$ and $D$ are defined as:
\begin{equation}\label{nn1}
A=N\cdot\mbox{SNR}_{1}\alpha,
\end{equation}
\begin{equation}\label{nn2}
B=N\cdot\mbox{SNR}_{2}\alpha,
\end{equation}
\begin{equation}\label{nn3}
C=N\sqrt{\mbox{SNR}_{1} \cdot \mbox{SNR}_{2}}\bar{\rho}\alpha,
\end{equation}and
\begin{equation}\label{nn4}
D=N\sqrt{\mbox{SNR}_{1}\cdot\mbox{SNR}_{2}}\tilde{\rho}\beta.
\end{equation}Thus our task of finding the expression of $\delta$ has been brought down to finding the root of (\ref{n7}).

\subsection{Expression of $\delta$  for different correlation factors}
The solution of (\ref{n7}), depending on different values of $\rho$, falls into the following three cases:
\begin{description}
\item[Case 1.] \emph{Non-zero imaginary part of the correlation coefficient $\rho$ ($\tilde{\rho}\neq0$)}: in this case (\ref{n7}) remains a quartic function in $\delta$. We know from the parameter transformation property of the CRB (cf. \cite{Kay93}, p.37) that $\mbox{CRB}(\delta)=\mbox{CRB}(-\delta)$, Thus, if $\delta$ is a root of (\ref{wc3}) (hence of (\ref{n7})), then $-\delta$ will also be a root thereof, which allows us to remove all the terms of odd degrees in (\ref{n7}), leading to a quadratic equation of $\delta^{2}$. Hence
\begin{equation}\label{n8}
D^{2}\delta^{4}+(C^{2}-AB)\delta^{2}+\frac{A+B}{2}+C=0.
\end{equation}The root of (\ref{n8}) is\footnote{The other root of (\ref{n8}), which is very large, is in contradiction with the observation made in Footnote 3, thus is regarded as a trivial solution and rejected.}:
\begin{equation}\label{nn5}
\begin{aligned}
\delta^{2}&=\frac{AB-C^{2}-\sqrt{(C^2-AB)^{2}-4D^{2}(\frac{A+B}{2}+C)}}{2D^{2}}\\
&=\frac{\gamma}{\kappa}\left(1-\sqrt{1-\frac{\alpha\kappa\phi}{\gamma^2}}\right),
\end{aligned}
\end{equation}where $\gamma=(1-\bar{\rho}^{2})\alpha^{2}$, $\kappa=2\tilde{\rho}^{2}\beta^{2}$ and
\begin{equation}\label{lt1}
\phi=\frac{1}{N}\left(\frac{1}{\mbox{SNR}_{1}}+\frac{1}{\mbox{SNR}_{2}}+\frac{2\bar{\rho}}{\sqrt{\mbox{SNR}_{1}\cdot\mbox{SNR}_{2}}}\right).
\end{equation}The existence of $\delta^{2}$ in (\ref{nn5}) is assured since under realistic conditions $(\alpha\kappa\phi/\gamma^2)\ll1$. Thus the ARL is given by: \begin{equation}\label{tl2}
\delta=\sqrt{\frac{\gamma}{\kappa}\left(1-\sqrt{1-\frac{\alpha\kappa\phi}{\gamma^2}}\right)}.
\end{equation}
\item[Case 2.] \emph{Not fully correlated signals with zero imaginary part of the correlation coefficient $\rho$ ($\tilde{\rho}=0$ and $\bar{\rho}\neq\pm1$)}: in this case $D=0$, $(C^{2}-AB)\neq0$, and (\ref{n7}) degenerates into $(C^{2}-AB)\delta^{2}+\frac{A+B}{2}+C=0$. Taking its positive root we have: \begin{equation}\label{tl}
\delta=\sqrt{\frac{\frac{A+B}{2}+C}{AB-C^2}}=\sqrt{\frac{\phi\alpha}{2\gamma}}.
\end{equation}The existence of $\delta$ is guaranteed from the fact that in this case both $\phi$ and $\gamma$ are greater than zero. It is worth noticing that an important special case of Case 2, in which both $\tilde{\rho}$ and $\bar{\rho}$ equal zero, namely, the two signals are uncorrelated, reduces (\ref{tl}) to
\begin{equation}
\delta=\sqrt{\frac{1}{2N\alpha}\left(\frac{1}{\text{SNR}_{1}}+\frac{1}{\text{SNR}_{2}}\right)}.
\end{equation}

\item[Case 3.] \emph{Fully correlated signals with zero imaginary part of the correlation coefficient $\rho$ ($\tilde{\rho}=0$ and $\bar{\rho}=\pm1$)}: in this case (\ref{n7}) degenerates to $\frac{A+B}{2}+C=0$ and a solution can not be found.\footnote{One can expect that for the case in which $\rho=\pm1$, i.e., the two signals are linearly dependent, the approximation made using a first order Taylor expansion is not sufficiently tight w.r.t. the true model. Thus in this case it entails a higher order Taylor expansion and thereby involves solving a sextic equation, the detailed analysis of which, unfortunately, is due to the space limitation beyond the scope of this paper.}
\end{description}

Now, combining the results of all three cases presented above yields our final expression of the ARL that can be written as:
\begin{align}\label{xg1}
\delta=\left\{
\begin{array}{>{\displaystyle}l>{\displaystyle}l}
\sqrt{\frac{\gamma}{\kappa}\left(1-\sqrt{1-\frac{\alpha\kappa\phi}{\gamma^2}}\right)}, \quad&\mbox{for}\ \tilde{\rho}\neq0\\
\sqrt{\frac{\phi\alpha}{2\gamma}}, \quad&\mbox{for}\ \tilde{\rho}=0\ \mbox{and}\ \bar{\rho}\neq\pm1\\
\mbox{(no closed-form expression available)}, \quad&\mbox{for}\ \tilde{\rho}=0\ \mbox{and}\ \bar{\rho}=\pm1
\end{array}
\right.
\end{align}Note that, for the uniform linear array (ULA) configuration the parameters in (\ref{xg1}) can be derived as $\alpha=\frac{M(M-1)(2M-1)}{6}d$ and $\beta=\frac{M^{2}(M-1)^{2}}{4}d$, where $d$ denotes the inter-sensor spacing.

\section{Simulations and Numerical Analysis}
\begin{figure}[htb]

  \centering
  \centerline{\includegraphics[width=9cm]{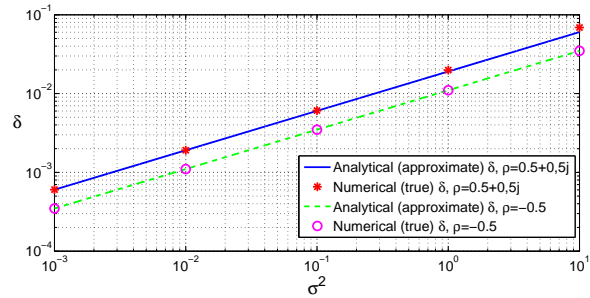}}

\caption{Numerical and analytical $\delta$ vs. $\sigma^2$ for $\varepsilon_1=\varepsilon_2=1$, with $\rho=0.5+0.5j$ and $\rho=-0.5$ respectively.}
\label{fig:1}
\end{figure}

\begin{figure}[htb]

  \centering
  \centerline{\includegraphics[width=9cm]{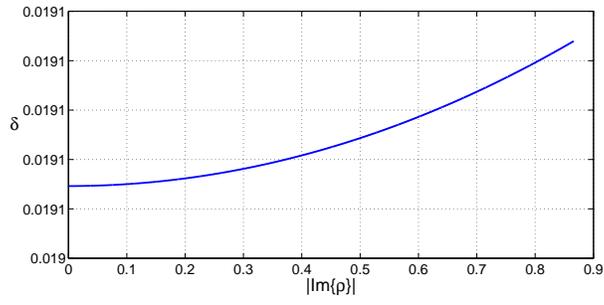}}

\caption{$\delta$ vs. $|\tilde{\rho}|$ for $\bar{\rho}=0.5$, $\varepsilon_1=\varepsilon_2=1$ and $\sigma^{2}=1$.}
\label{fig:2}
\end{figure}

\begin{figure}[htb]

  \centering
  \centerline{\includegraphics[width=9cm]{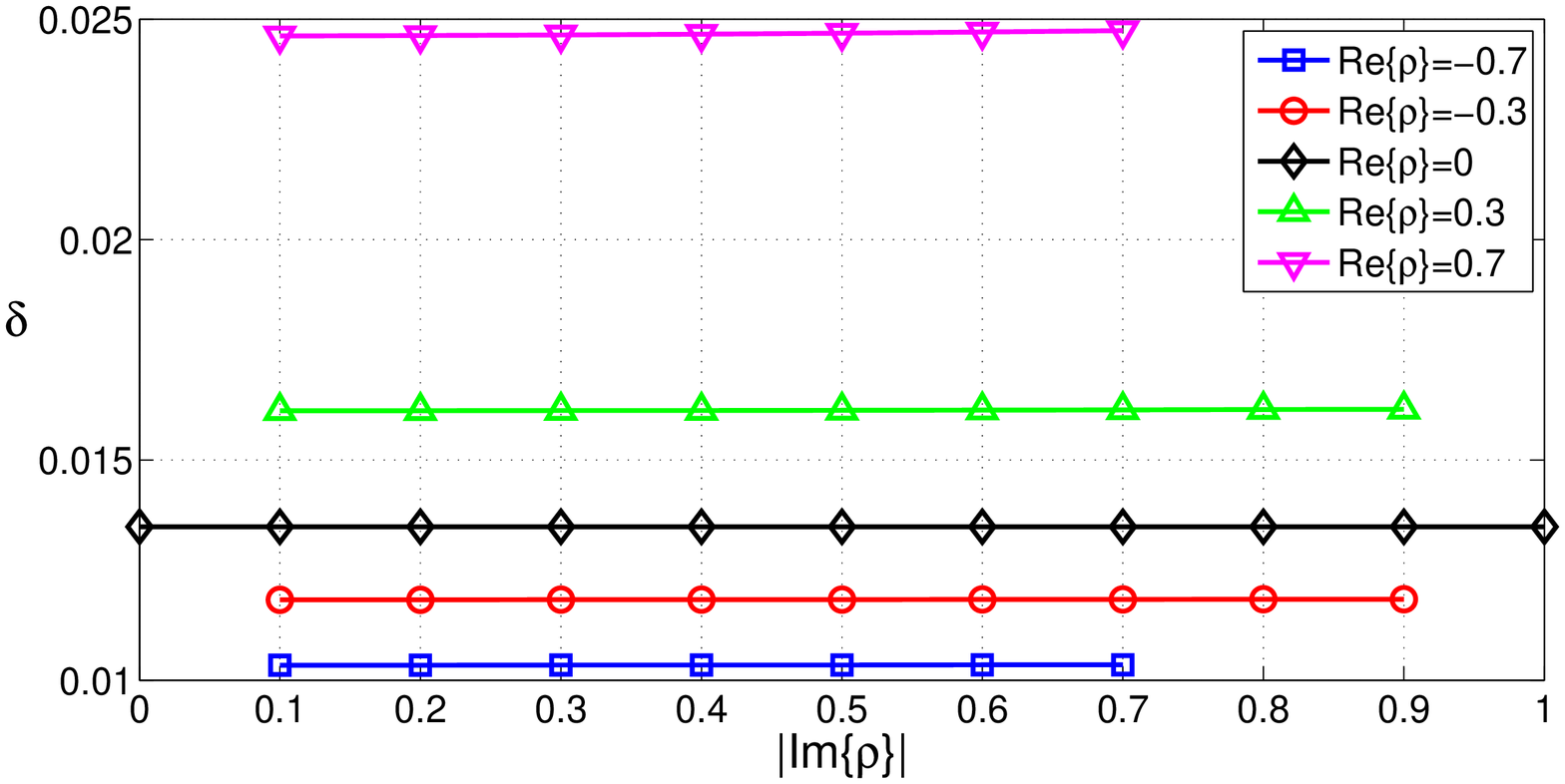}}

\caption{$\delta$ vs. $|\tilde{\rho}|$ for $\varepsilon_1=\varepsilon_2=1$, $\sigma^{2}=1$ and various $\bar{\rho}$.}
\label{fig:3}
\end{figure}

\begin{figure}[htb]

  \centering
  \centerline{\includegraphics[width=9cm]{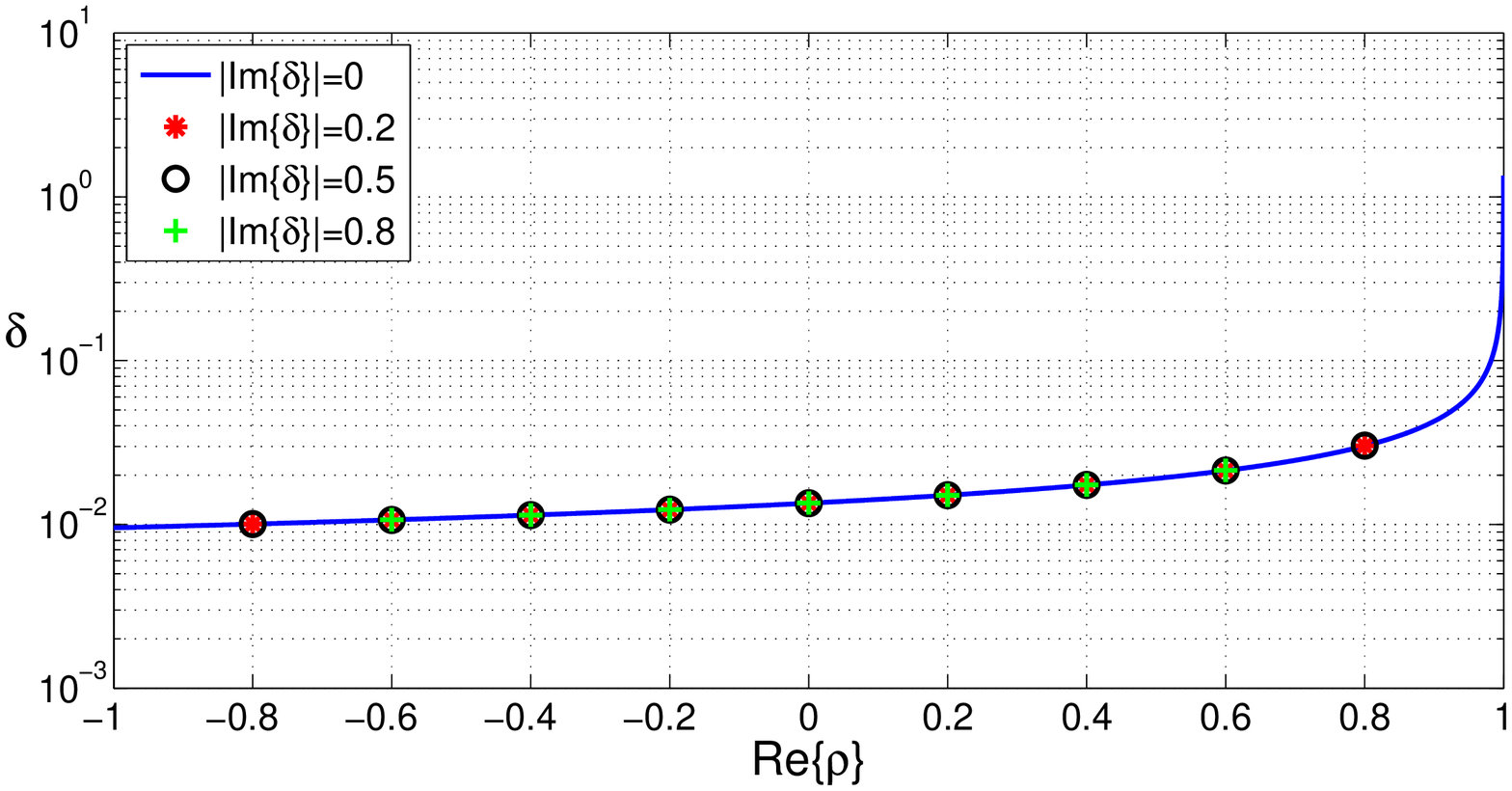}}

\caption{$\delta$ vs. $\bar{\rho}$ for $\varepsilon_1=\varepsilon_2=1$, $\sigma^{2}=1$ and various $\tilde{\rho}$.}
\label{fig:4}
\end{figure}

\begin{figure}[!ht]

  \centering
  \centerline{\includegraphics[width=9cm]{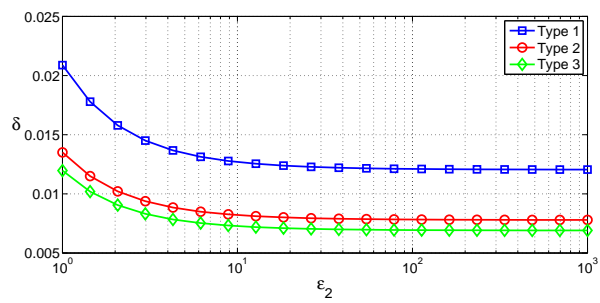}}

\caption{$\delta$ vs. $\varepsilon_2$ for $\rho=0.5+0.5j$, $\varepsilon_1=1$, $\sigma^{2}=1$ and various array configurations.}
\label{fig:5}
\end{figure}

The context of our simulation is a ULA of $M=6$ sensors with half-wave length inter-element spacing. The snapshot number is given by $N=100$. Our results are as follows:
\begin{itemize}
\item In Fig.~\ref{fig:1} we validate our approximate analytical expression of $\delta$ in (\ref{xg1}) for two cases ($\tilde{\rho}\neq0$; $\tilde{\rho}=0 \& \bar{\rho}\neq1$) by comparing it with the true $\delta$ (obtained by solving (\ref{wc3}) numerically) and show that both results are identical.
\item As is revealed by (\ref{xg1}), the concrete waveforms of the signals has no effect on $\delta$, which only depends on the two signal's respective strengths ($\varepsilon_1$, $\varepsilon_2$) and the correlation $\rho$ between them. Furthermore, note that either $\bar{\rho}$ or $\tilde{\rho}$ plays its role separately. Fig.~\ref{fig:2} shows that, with a fixed $\bar{\rho}$, the ARL $\delta$ slightly increases with the value of $|\tilde{\rho}|$. However, this impact is so limited compared to that of the parameter $\bar{\rho}$, such that the former is practically negligible, (cf. Fig.~\ref{fig:3} and Fig.~\ref{fig:4}, both of which show that $\delta$ increases notably as $\bar{\rho}$ raises, while remains nearly unaltered with the change of $\tilde{\rho}$.) This fact can be explained by considering, for $(\alpha\kappa\phi/\gamma^2)\ll1$, the first order Taylor expansion to (\ref{tl2}) around $(\alpha\kappa\phi/\gamma^2)=0$ that is given by:
    \begin{equation}
    \delta\approx\sqrt{\frac{\gamma}{\kappa}\left(1-\left(1-\frac{\alpha\kappa\phi}{2\gamma^2}\right)\right)}=\sqrt{\frac{\phi\alpha}{2\gamma}},
    \end{equation}which is the same expression as (\ref{tl}), independent of $\tilde{\rho}$.

\item The dependence of $\delta$ on the signal strengths is reflected in the expression of $\phi$ (cf. (\ref{lt1})), where we see that if the strength of one signal is much greater than the other, e.g., $\varepsilon_1\gg\varepsilon_2$, then $\phi\approx1/(N\cdot\text{SNR}_{2})$, and $\delta$ becomes restricted by the weaker signal. Thus enhancing the strength of only one signal cannot infinitely diminish $\delta$, as is shown by Fig.~\ref{fig:5}, in which we increase $\varepsilon_1$ from $1$ to $1000$ while keeping $\varepsilon_2=1$, and find that $\delta$ converges to a certain value (determined by $\varepsilon_2$).
\item Fig.~\ref{fig:5} also investigates the impact of the sensor array geometry on $\delta$ (cf. Table 1) and reveals that a loss of sensors in the array configuration has a considerable impact on $\delta$ only when it causes a diminution of the aperture size of the array, as in the case of Type 1. If, however, the array aperture remains unchanged, as in the case of Type 2, this impact is considerably mitigated.
\end{itemize}

\begin{table}[htb]
 \begin{center}
 \begin{tabular}{lc}
  \toprule
  Array Type & Geometric Configuration\\
  \midrule
  Type 1 & $\circ$ \ $\bullet$ \ $\bullet$ \ $\circ$ \ $\bullet$ \ $\bullet$ \ $\circ$ \ $\circ$\\
Type 2 & $\bullet$ \ $\circ$ \ $\circ$ \ $\circ$ \ $\circ$ \ $\bullet$ \ $\bullet$ \ $\bullet$\\
    Type 3 & $\bullet$ \ $\bullet$ \ $\bullet$ \ $\bullet$ \ $\bullet$ \ $\bullet$ \ $\bullet$ \ $\bullet$\\
  \bottomrule
 \end{tabular}
   \caption{Different array geometric configurations. $\bullet$ and $\circ$ represent the position of sensor
and missing sensors, respectively. The inter-element spacing is half-wave length.}
 \end{center}
\end{table}

%
%
%

\section{Summary}
In this paper we studied the angular resolution between two point sources and provide a closed-form expression for the ARL $\delta$, the validity of which was confirmed by simulation. We also noticed that $\delta$ is not dependent on the special waveforms of the signals, but only on their strengths and the correlation factor between them, and that the imaginary part of $\rho$ only has a negligible impact on $\delta$, while the impact of the real part of $\rho$ is decisive. Furthermore, we showed that $\delta$ is constrained by the weaker signal, and therefore cannot be infinitely decreased. Finally the impact of different array geometries on $\delta$ is discussed.


\clearpage

\bibliographystyle{IEEEbib}
\bibliography{zhxtc}

\begin{thebibliography}{10}

\bibitem{KV96}
H.~Krim and M.~Viberg,
\newblock ``Two decades of array signal processing research: the parametric
  approach,''
\newblock {\em {IEEE} Signal Processing Mag.}, vol. 13, no. 4, pp. 67--94,
  1996.

\bibitem{S05}
S.~T. Smith,
\newblock ``Statistical resolution limits and the complexified {C}ram\'{e}r
  {R}ao bound,''
\newblock {\em {IEEE} Trans. Signal Processing}, vol. 53, pp. 1597--1609, May
  2005.

\bibitem{ShaMil05}
M.~Shahram and P.~Milanfar,
\newblock ``On the resolvability of sinusoids with nearby frequencies in the
  presence of noise,''
\newblock {\em {IEEE} Trans. Signal Processing}, vol. 53, no. 7, pp.
  2579--2588, July 2005.

\bibitem{EBRM10}
M.~N. {El~Korso}, R.~Boyer, A.~Renaux, and S.~Marcos,
\newblock ``Statistical resolution limit for multiple parameters of interest
  and for multiple signals,''
\newblock in {\em Proc. ICASSP}, Dallas, TX, Mar. 2010, pp. 3602--3605.

\bibitem{C73}
H.~Cox,
\newblock ``Resolving power and sensitivity to mismatch of optimum array
  processors,''
\newblock {\em J. Acoust. Soc.}, vol. 54, no. 3, pp. 771--785, 1973.

\bibitem{SM04}
M.~Shahram and P.~Milanfar,
\newblock ``Imaging below the diffraction limit: {A} statistical analysis,''
\newblock {\em {IEEE} Trans. Image Processing}, vol. 13, no. 5, pp. 677--689,
  May 2004.

\bibitem{LN07}
Z.~Liu and A.~Nehorai,
\newblock ``Statistical angular resolution limit for point sources,''
\newblock {\em {IEEE} Trans. Signal Processing}, vol. 55, no. 11, pp.
  5521--5527, Nov. 2007.

\bibitem{L92}
H.~B. Lee,
\newblock ``The {Cram\'er-Rao} bound on frequency estimates of signals closely
  spaced in frequency,''
\newblock {\em {IEEE} Trans. Signal Processing}, vol. 40, no. 6, pp.
  1507--1517, 1992.

\bibitem{EBRM11a}
M.~N.~El Korso, R.~Boyer, A.~Renaux, and S.~Marcos,
\newblock ``Statistical resolution limit of the uniform linear cocentered
  orthogonal loop and dipole array,''
\newblock {\em IEEE Trans. Signal Processing}, vol. 59, no. 1, pp. 425--431,
  Jan. 2011.

\bibitem{L94}
H.~B. Lee,
\newblock ``The {Cram\'er-Rao} bound on frequency estimates of signals closely
  spaced in frequency (unconditional case),''
\newblock {\em {IEEE} Trans. Signal Processing}, vol. 42, no. 6, pp.
  1569--1572, 1994.

\bibitem{D98}
E.~Dilaveroglu,
\newblock ``Nonmatrix {Cram\'er-Rao} bound expressions for high-resolution
  frequency estimators,''
\newblock {\em {IEEE} Trans. Signal Processing}, vol. 46, no. 2, pp. 463--474,
  Feb. 1998.

\bibitem{AW08}
A.~Amar and A.J. Weiss,
\newblock ``Fundamental limitations on the resolution of deterministic
  signals,''
\newblock {\em {IEEE} Trans. Signal Processing}, vol. 56, no. 11, pp.
  5309--5318, Nov. 2008.

\bibitem{zhxtc6}
M.~N. {El~Korso}, R.~Boyer, A.~Renaux, and S.~Marcos,
\newblock ``Statistical resolution limit for source localization with clutter
  interference in a {MIMO} radar context,''
\newblock {\em IEEE Trans. Signal Processing}, vol. 60, no. 5, pp. 987--992,
  Feb. 2012.

\bibitem{zhxtc7}
D.~T. Vu, M.~N. {El~Korso}, R.~Boyer, A.~Renaux, and S.~Marcos,
\newblock ``Angular resolution limit for vector-sensor arrays: detection and
  information theory approaches,''
\newblock in {\em Proc. IEEE Workshop on Statistical Signal Processing,
  SSP-2011}, Nice, France, June 2011, pp. 9--12.

\bibitem{AD08}
H.~Abeida and J.-P. Delmas,
\newblock ``Statistical performance of {MUSIC}-like algorithms in resolving
  noncircular sources,''
\newblock {\em IEEE Trans. Signal Processing}, vol. 56, no. 6, pp. 4317--4329,
  Sept. 2008.

\bibitem{AJS09}
Y.~I. Abramovich, B.~A. Johnson, and N.~K. Spencer,
\newblock ``Statistical nonidentifiability of close emitters:
  {M}aximum-likelihood estimation breakdown,''
\newblock in {\em EUSIPCO}, Glasgow, Scotland, Aug. 2009.

\bibitem{zhxtc8}
R.~Boyer,
\newblock ``Performance bounds and angular resolution limit for the moving
  co-located {MIMO} radar,''
\newblock {\em IEEE Trans. Signal Processing}, vol. 59, no. 4, pp. 1539--1552,
  Apr. 2011.

\bibitem{God97}
L.C. Godara,
\newblock ``Applications of antenna arrays to mobile communications: {II.}
  {B}eam-forming and direction of arrival considerations,''
\newblock {\em {IEEE} Trans. Antennas Propagat.}, vol. 85, no. 8, pp.
  1195--1245, Aug. 1997.

\bibitem{LSZ96}
J.~Li, P.~Stoica, and D.~Zheng,
\newblock ``Efficient direction and polarization estimation with a cold
  array,''
\newblock {\em {IEEE} Trans. Antennas Propagat.}, vol. 44, no. 4, pp. 539--547,
  Apr. 1996.

\bibitem{zhxtc3}
F.~R{\"o}mer and M.~Haardt,
\newblock ``Deterministic {C}ram\'er-{R}ao bounds for strict sense non-circular
  sources,''
\newblock in {\em Proc. ITG/IEEE Workshop on Smart Antennas (WSA'07)}, Vienna,
  Austria, Feb. 2007.

\bibitem{SM05}
P.~Stoica and R.L. Moses,
\newblock {\em Spectral Analysis of Signals},
\newblock Prentice Hall, NJ, 2005.

\bibitem{leh83}
E.~L. Lehmann,
\newblock {\em Theory of Point Estimation},
\newblock Wiley, New York, 1983.

\bibitem{SN89}
P.~Stoica and A.~Nehorai,
\newblock ``{MUSIC}, maximum likelihood and the {C}ram\'{e}r {R}ao bound,''
\newblock {\em {IEEE} Trans. Acoust., Speech, Signal Processing}, vol. 37, pp.
  720--741, May 1989.

\bibitem{zhxtc9}
M.~Kaveh,
\newblock ``Performance bounds and angular resolution limit for the moving
  co-located {MIMO} radar,''
\newblock {\em {IEEE} Trans. Acoust., Speech, Signal Processing}, vol. 34, no.
  2, pp. 331--341, Apr. 1986.

\bibitem{Kay93}
S.~M. Kay,
\newblock {\em Fundamentals of Statistical Signal Processing : {E}stimation
  Theory}, vol.~1,
\newblock Prentice Hall, NJ, 1993.

\end{thebibliography}

\end{document}